\documentclass[letterpaper,conference]{IEEEtran}
\IEEEoverridecommandlockouts

\usepackage{cite}
\usepackage{amsmath,amssymb,amsfonts}
\usepackage{algorithmic}
\usepackage{graphicx}
\usepackage{textcomp}
\usepackage{subfigure}
\usepackage{epstopdf}
\usepackage{multicol}
\usepackage{multirow}
\usepackage{siunitx}
\usepackage{float}
\usepackage{footnote}
\usepackage[table,xcdraw]{xcolor}
\usepackage[margin = 0.625in,top=0.75in,bottom=1in]{geometry}
\usepackage{draftwatermark}
\SetWatermarkText{Camera Ready}
\SetWatermarkScale{0.5}

\def\BibTeX{{\rm B\kern-.05em{\sc i\kern-.025em b}\kern-.08em
		T\kern-.1667em\lower.7ex\hbox{E}\kern-.125emX}}

\DeclareMathOperator*{\argmax}{arg\,max}

\begin{document}
	
	\title{Classifying the Order of Higher Derivative Gaussian Pulses in Terahertz Wireless Communications
		\vspace{-6mm}
	}
	\author{\IEEEauthorblockN{Shree Prasad M., Trilochan Panigrahi}
		\IEEEauthorblockA{\textit{Dept. of Electronics and Communication Engg.} \\ \textit{National Institute of Technology Goa},
			Goa, India \\
			shreeprasadm@gmail.com, tpanigrahi@nitgoa.ac.in}
		\vspace{-13.761mm}
		\and
		\and
		\IEEEauthorblockN{Mahbub Hassan}
		\IEEEauthorblockA{\textit{School of Computer Science and Engineering} \\ \textit{University of New South Wales},
			Sydney, Australia \\
			mahbub@cse.unsw.edu.au}
		\vspace{-13.761mm}
	}
		\maketitle
		\begin{abstract}
			The terahertz band is considered the last frontier for wireless communications and expected to play a significant role in beyond 5G networks. Besides supporting extremely high data rates for existing devices, the terahertz band is also expected to connect future nanoscale devices using graphene-based nano-antenna, which happens to radiate in 0.1-10 THz band. In this band, higher order derivatives of Gaussian pulses can provide energy-efficient communication for nanodevices.  In this paper, we propose a metric, called root mean square (RMS) frequency spread, to detect the derivative order of the pulse at the receiving base station using uniform linear array antennas. Simulation experiments demonstrate that RMS frequency spread can be used to detect the derivative order of Gaussian pulses with 99\% accuracy for distances up to 50 cm. This finding opens up a new design space for nanoscale terahertz communication, which can encode information in time derivative order of the transmitted pulse.	
		\end{abstract}
		\maketitle
		\vspace{-3mm}
		\section{Introduction}
		\vspace{-1mm}
		While the millimeter wave band between 30-300GHz \cite{b1_1} is already supporting 5G developments, the terahertz band extending up to 10 THz has been earmarked as the last trump card to meet the future needs beyond 5G. Besides supporting extremely high data rates for existing devices, the terahertz band is also expected to connect future nanoscale devices using graphene-based nano-antenna, which happens to radiate in 0.1-10 THz band \cite{b1}. Nanoscale devices, such as tiny sensors measuring only a few hundred nanometers, are already being designed, fabricated, and tested in the laboratories. Powered by nanostructured design, these nanosensors are capable of detecting the smallest changes in physical variables, such as pressure, vibrations, temperature, and concentrations in chemical and biological molecules. Such nanoscale event detection and wireless communication will open up new Internet of Things (IoT) capabilities for gathering knowledge at an unprecedented depth and scale, offering massive improvements in healthcare, agriculture, transportation, security, surveillance, industrial chemistry, and so on. Researcher are now pursuing this new direction of IoT under the banner of Internet of Nano Things (IoNT) \cite{b2} with nanoscale monitoring techniques explored for human body \cite{b3,b4}, plants \cite{b5}, chemical processes \cite{b6}, and so on. 
		
		Graphene-based nanoantenna is a significant step forward for realizing the vision of IoNT, but sustained event monitoring remains a major challenge due to extremely limited energy supply at the nanoscale. To address the energy issue at the nanoscale, pulse-based communication protocols are being developed  \cite{b9} where all data is transmitted as a series of short (a few hundred femtoseconds) higher time derivative Gaussian pulses. Use of such short pulses reduces the total energy consumption drastically compared to conventional \textit{continuous wave} wireless communications. 
		
		In this paper, we seek to detect mechanisms for the receiving base station to detect the derivative order of the transmitted pulse. The motivation for this is to open up the potential for encoding information in the time derivative of the pulse, which could increase resource utilization or decrease overall energy consumption for nanoscale wireless communications. For example, a nanoscale device could simply switch between two derivative orders to update some binary status of an event that it is monitoring. Similarly, different nanodevices can be configured with different derivative orders for implicit device identification. Many other new protocol designs can be envisaged by exploiting the derivative order detection capability at the base station. 
		
	    Detecting the derivative order of Gaussian pulses is a challenging problem, which to our knowledge has not been addressed in the literature. We observe that different derivative orders have different pulse durations. In other words, increasing the time derivative order increases pulse durations, which consequently reduces the pulse bandwidth as illustrated in Fig. \ref{fig:HPLS}. For order detection, we can, therefore, exploit only two features, the pulse duration or the pulse bandwidth. Detecting pulse duration for femtosecond pulses is impractical with current state-of-the-art measurement techniques. Unfortunately, as will be shown later in Section III, the complex propagation dynamics in the terahertz, which are significantly influenced by frequency selective molecular absorption laws, make it equally impractical to estimate pulse bandwidth at the receiver.
		
		In this paper, we propose a novel pulse bandwidth estimation metric, called root mean square (RMS) frequency spread, which enables us to overcome the molecular absorption related challenges and accurately detect bandwidth differences between different derivative pulses. We achieve this by using uniform linear array (ULA) antennas at the base station, which can afford more complexity compared to transmitting nanosensors.  
		
		The contributions in this paper can be summarized as follows:\\
	$\bullet$ We define RMS frequency spread and design its implementation using ULA to detect the derivative order of Gaussian pulses for the first time.\\ 
	$\bullet$ Using simulation, we demonstrate that the proposed method can detect derivative order with 99\% accuracy from a distance of 50 cm for snapshot observation interval greater than 10 ps. We achieve this performance by observing and analyzing a \textit{single} pulse from a \textit{single} snapshot.\\
	$\bullet$ Our study reveals that the order classification accuracy improves with increasing duration of snapshot observation. In contrast, classification accuracy cannot be increased by increasing the number of antenna elements due to the use of a \textit{single} snapshot.
	
		The rest of the paper is structured as follows. Higher time derivative order Gaussian pulses are reviewed in section II, followed by the terahertz channel response and molecular absorption laws. We present the system model in section IV. In section V, simulations experiments are presented and discussed. Section VI concludes the paper.
		\vspace{-2mm}
		\section{Higher time derivative order Gaussian pulses}
\begin{figure}[t]
			\centering
			\subfigure{
				\includegraphics[width=0.48\columnwidth, height = 3cm]{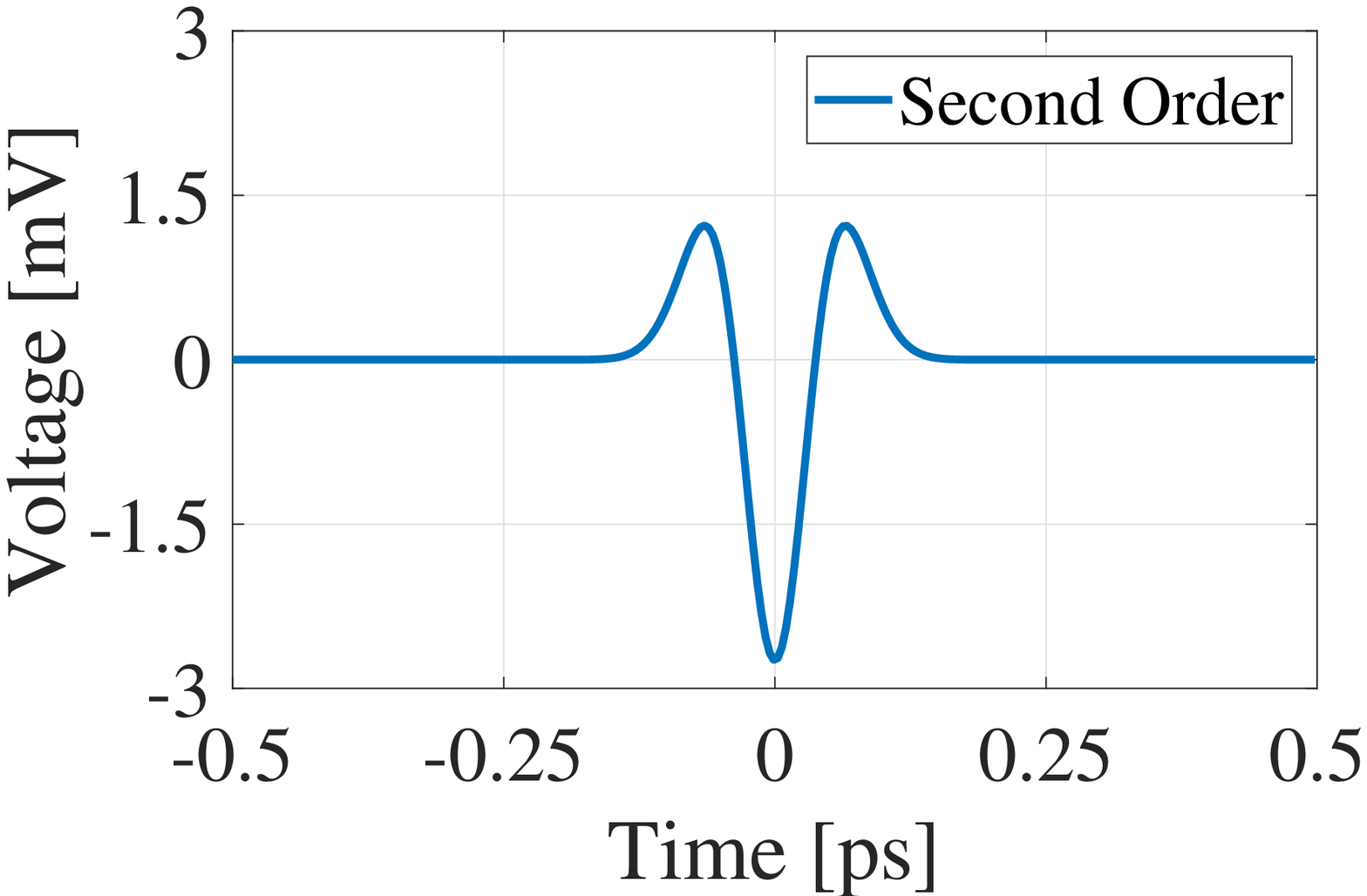}
			}
			\vspace{-2mm}
			\hspace{-5mm}
			\subfigure{
				\includegraphics[width=0.48\columnwidth, height = 3cm]{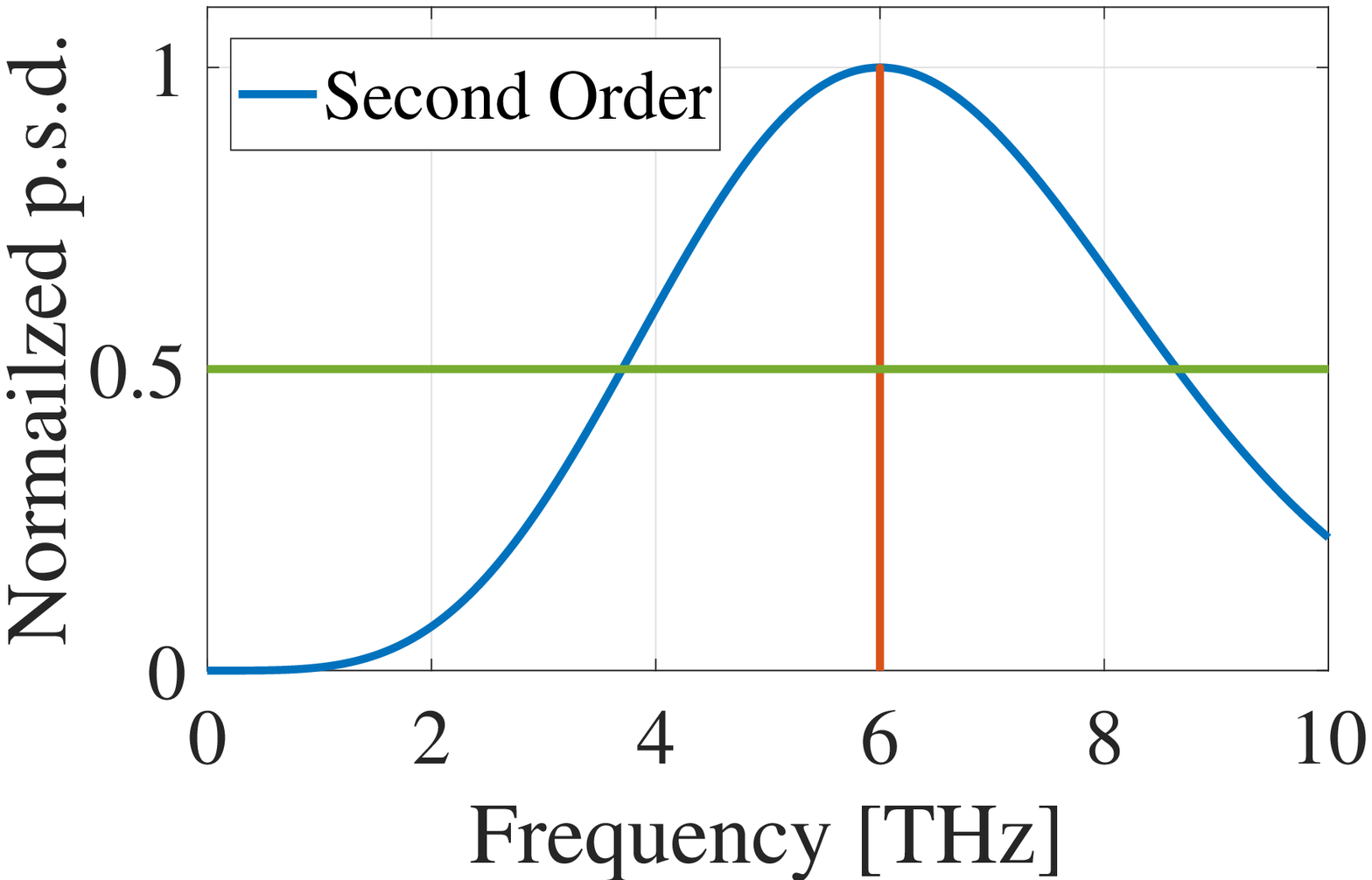}
			}
			\vspace{-2mm}
			\subfigure{
				\includegraphics[width=0.48\columnwidth, height = 3cm]{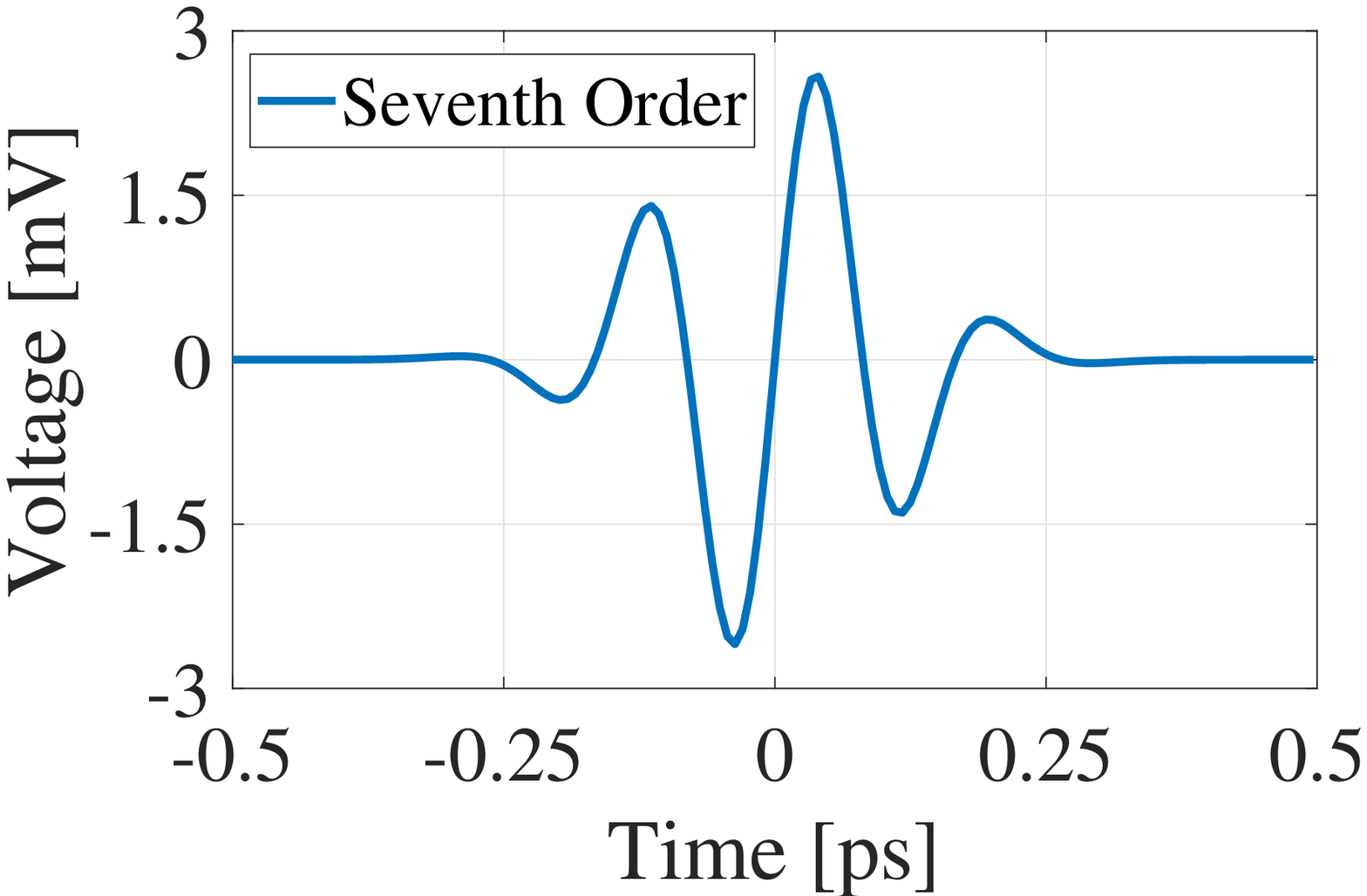}
			}
			\vspace{-1mm}
			\hspace{-5mm}
			\subfigure{
				\includegraphics[width=0.48\columnwidth, height = 3cm]{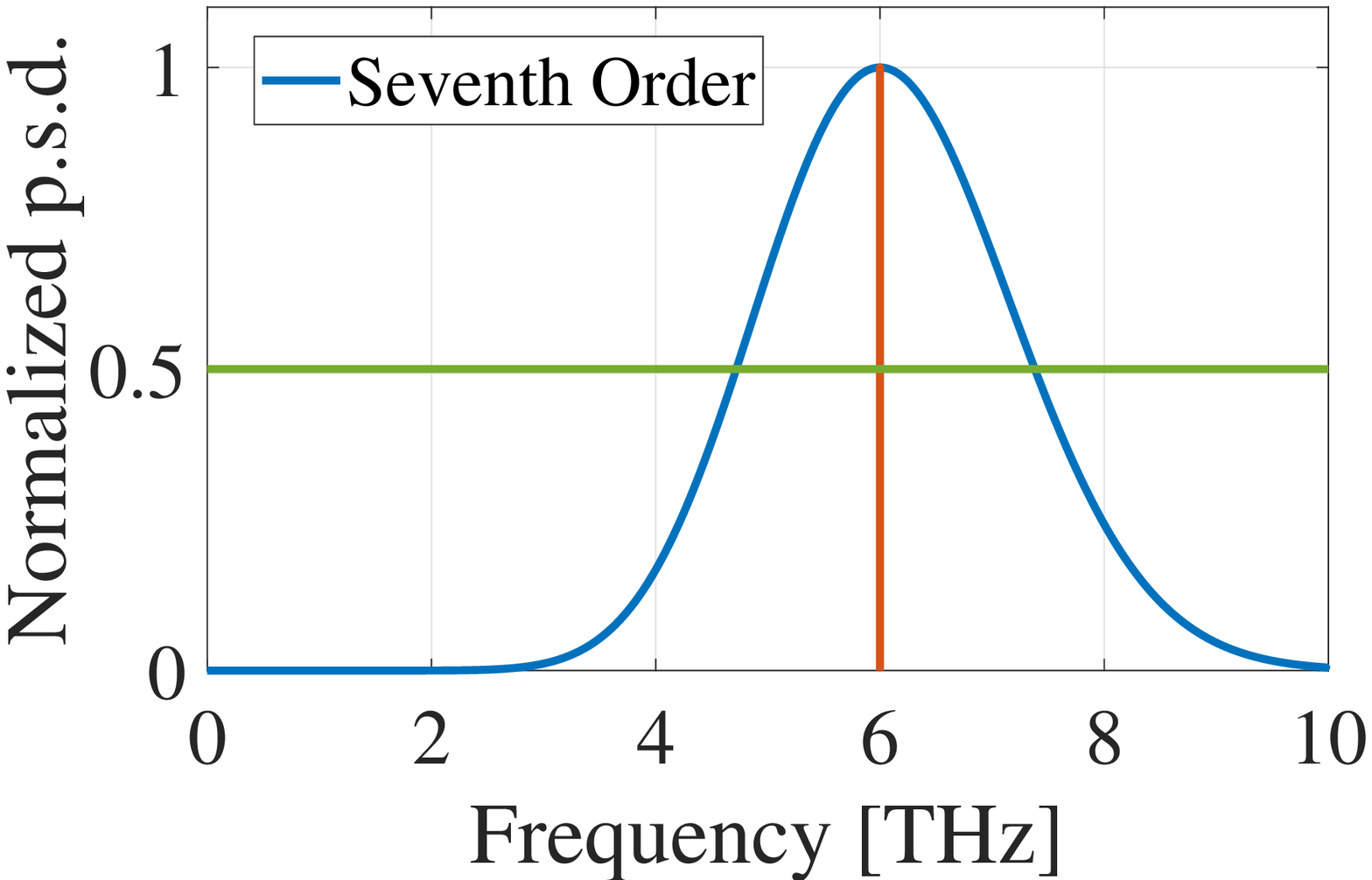}
			}
			\vspace{-1mm}
			\subfigure{
				\includegraphics[width=0.48\columnwidth, height = 3cm]{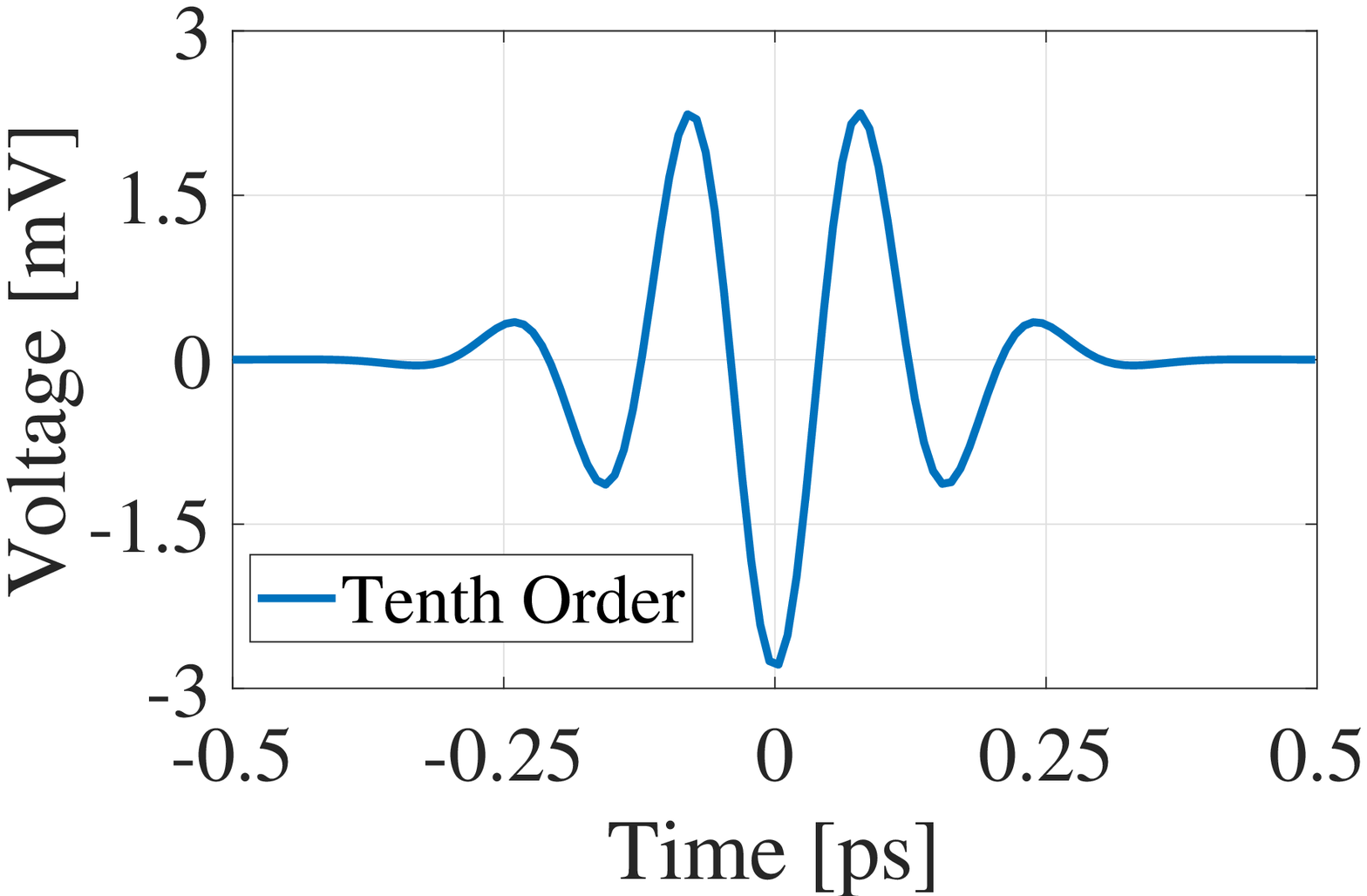}
			}
			\hspace{-5mm}
			\subfigure{
				\includegraphics[width=0.48\columnwidth, height = 3cm]{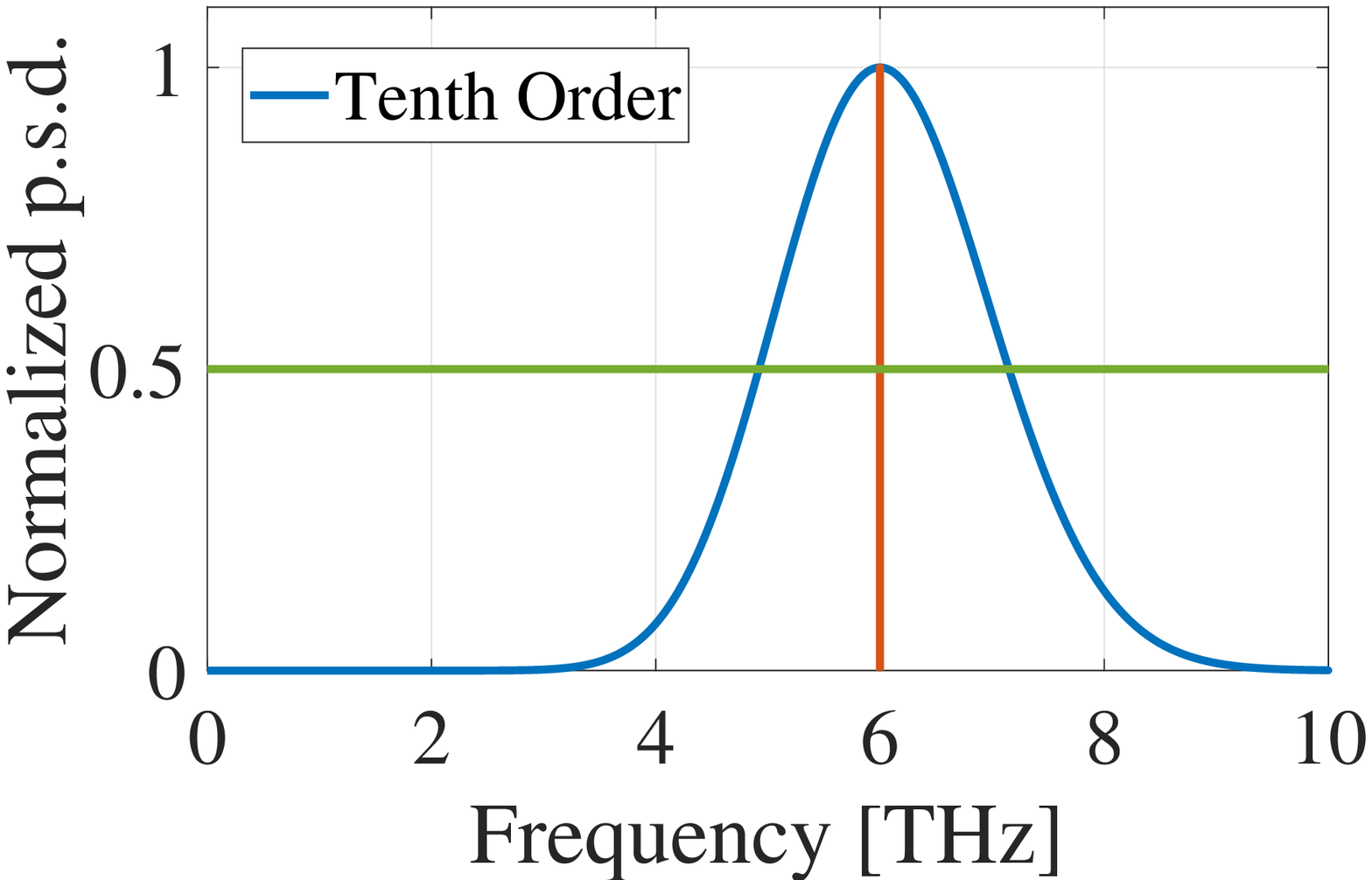}
			}
			\vspace{-2mm}
			\caption{ Higher order time derivative Gaussian pulses of 1\si{\micro} W power with 6 THz center frequency. Increasing the time derivative order increases pulse durations in the time domain (left column), which consequently reduces pulse bandwidth in the frequency domain (right column).}
			\label{fig:HPLS}
			\vspace{-4mm}
		\end{figure}

		\vspace{-1mm}
		 The Fourier representation of time derivative of the Gaussian pulse with the order of derivative $n$ is also Gaussian shaped and is represented as\cite{b9}
		\vspace{-1mm}
		\begin{equation}
		P_{n}\left( f\right)  = a_{n}\left(j 2\pi f\right)^{n}  e^{-0.5\left( 2\pi\sigma f\right)^{2} }
	\vspace{-1mm}
		\end{equation}
	where $a_{n}$ is the normalizing constant to adjust the pulse energy, and $\sigma$ is the standard deviation of the Gaussian pulse in seconds. Pulse duration is defined as the multiple of the standard deviation, which contains 99.99\% of the pulse energy. The derivative order is related to the standard deviation according to the following rule\cite{b12}
		\vspace{-1.5mm}
		\begin{equation}\label{eq:cent_freq}
		f_{c} = \frac{\sqrt{n}}{2\pi \sigma}
		\vspace{-1.5mm}
		\end{equation} 
		where $f_c$ denotes the center frequency of power spectral density (p.s.d.) of the Gaussian pulse. It is clear that, for a given center frequency, an increase in derivative order will also increase the standard deviation (and hence pulse duration) in the time domain, which will consequently reduce its bandwidth in the frequency domain as illustrated in Fig. \ref{fig:HPLS}. 
		\begin{table*}
			\centering
			\caption{Center frequency, half power frequency, and RMS frequency spread of different Order Gaussian pulses.}
			\vspace{-3mm}
			\label{tab:TABLE1}
			\resizebox{\textwidth}{0.08\textwidth}{\begin{tabular}{|c|c|c|c|c|c|c|c|c|c|c|c|c|c|c|c|c|c|}
					\hline
					\multirow{2}{*}{Order} & \multicolumn{4}{c|}{$f_{c}$ = 3 THz} & \multicolumn{4}{c|}{$f_{c}$ = 6 THz} & \multirow{2}{*}{Order} & \multicolumn{4}{c|}{$f_{c}$ = 3 THz} & \multicolumn{4}{c|}{$f_{c}$ = 6 THz} \\ \cline{2-9} \cline{11-18} 
					&$f_{l}\;\text{[THz]}$      & $f_{h}\;\text{[THz]}$ & $B_{3\:dB}\;\text{[THz]}$  & $\Gamma_{n,f_{c}}\;\text{[THz]}$ & $f_{l}\;\text{[THz]}$   & $f_{h}\;\text{[THz]}$   &$B_{3\:dB}\;\text{[THz]}$ & $\Gamma_{n,f_{c}}\;\text{[THz]}$   &                        & $f_{l}\;\text{[THz]}$    & $f_{h}\;\text{[THz]}$ & $B_{3\:dB}\;\text{[THz]}$ & $\Gamma_{n,f_{c}}\;\text{[THz]}$ & $f_{l}\;\text{[THz]}$     & $f_{h}\;\text{[THz]}$& $B_{3\:dB}\;\text{[THz]}$ & $\Gamma_{n,f_{c}}\;\text{[THz]}$  \\ \hline
				1                      & 1.444  & 4.909  & 3.464 & 1.451 & 2.889  & 9.819  & 6.929 & 2.119 & 6                      & 2.310  & 3.747 & 1.437 & 0.609 & 4.620  & 7.495 & 2.874 & 1.207 \\ \hline
					2                      & 1.85   & 4.324  & 2.473 & 1.038 & 3.701  & 8.649  & 4.947 & 1.809 & 7                      & 2.359  & 3.690 & 1.331 & 0.564 & 4.718  & 7.381 & 2.662 & 1.124 \\ \hline
					3                      & 2.045  & 4.071  & 2.026 & 0.855 & 4.090  & 8.142  & 4.052 & 1.597 & 8                      & 2.398  & 3.644 & 1.245 & 0.528 & 4.797  & 7.289 & 2.491 & 1.054 \\ \hline
					4                      & 2.164  & 3.922  & 1.757 & 0.744 & 4.329  & 7.844  & 3.515 & 1.436 & 9                      & 2.431  & 3.606 & 1.174 & 0.498 & 4.863  & 7.213 & 2.349 & 0.995 \\ \hline
					5                      & 2.248  & 3.821  & 1.573 & 0.666 & 4.496  & 7.643  & 3.147 & 1.309 & 10                     & 2.459  & 3.574 & 1.114 & 0.472 & 4.919  & 7.149 & 2.229 & 0.945 \\ \hline
			\end{tabular}}
		\vspace{-5mm}
		\end{table*}
	\vspace{-2mm}
		\section{Terahertz Channel}
		\vspace{-0.15cm}
		The channel response and ambient noise affecting the propagation of electromagnetic waves in terahertz band channel is modeled using radiative transfer theory\cite{b9} and is reviewed in this section.
		\subsection{Terahertz Channel Impulse Response}
		The terahertz channel response accounts for both spreading loss and molecular absorption loss and is represented in the frequency domain as 
		\vspace{-1mm}
		\begin{equation}\label{eq:chresp}
		H\left( f,d_{r}\right) = H_{spread}\left( f,d_{r}\right) H_{abs}\left( f,d_{r}\right)  
		\vspace{-1mm}
		\end{equation}
		where $H_{spread}\left(f,d_{r} \right) $ and $H_{abs}\left(f,d_{r} \right) $ represents spreading loss and molecular absorption loss, and are given by
		\begin{equation}
		\resizebox{0.75\columnwidth}{!}{$H_{spread}\left( f,d_{r}\right) =\left(  \frac{c_{o}}{4\pi d_{r} f_{o}}\right) \exp\left( -\frac{j 2\pi f d_{r}}{c_{o}}\right)$} 
		\vspace{-1mm}
		\end{equation}
		\begin{equation}
		H_{abs}\left( f,d_{r}\right) = \exp(-0.5 k\left(f \right)d_{r} ) 
		\end{equation}
		where $f$ denotes frequency, $c_{o}$ is the velocity of light in vacuum, $f_{o}$ is the center frequency of the graphene antenna, $d_{r}$ is the path length, and $k\left( f\right) $ is the medium absorption coefficient. The medium absorption coefficient $k\left( f\right) $ of the terahertz channel at frequency $f$ composed of $Q$ type molecules is given as
		\vspace{-0.5mm}
		\begin{equation}
		\vspace{-0.25cm}
		\resizebox{0.4\columnwidth}{!}{$k\left( f\right)  = \sum\limits_{q=1}^{Q}x_{q}K_{q}\left( f\right).$}
		\end{equation} 
		where $x_{q}$ is the mole fraction of molecule type $q$ and $K_{q}$ is the absorption coefficient of individual molecular species.
		\vspace{-2mm}
		\subsection{Molecular Absorption Noise}
\begin{figure}
				\vspace{-2mm}
			\centering
			\includegraphics[width=0.8\columnwidth, height =3.5cm]{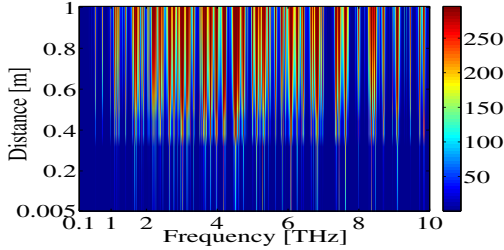}
			\vspace{-3mm}
			\caption{Molecular noise temperature in Kelvin as a function of frequency and path length in standard air medium with 1.86\%  of water vapor.}
			\label{fig:TEMP_GRAP}
		\end{figure}

		\vspace{-0.18cm}
		   The ambient noise in terahertz channel is the molecular absorption noise, which arises only during transmission of pulses. For transceivers operating in the terahertz band, which are fabricated using graphene material, the effect of thermal noise is negligible\cite{b10}. The molecular noise temperature $T_{mol}$ created due to transmission of pulses is given in \eqref{eqn:noitemp}. Fig. \ref{fig:TEMP_GRAP} shows the molecular noise temperature as a function of frequency and path length for standard air medium with 1.86 \% concentration of water vapor molecules\cite{b11}. It is observed, that molecular absorption noise becomes significant for path length above 0.4 m. It is also observed that molecular absorption noise is insignificant in certain frequency bands even for path length above 0.4 m. The total molecular absorption noise p.s.d. $S_{N}\left( f,d_{r}\right) $ affecting the transmission of pulse is the sum of background atmospheric noise p.s.d \(S_{N_{B}}\left( f,d_{r}\right)\) and the self induced noise p.s.d. $S_{N_{P}}\left(f,d_{r} \right)$ and is given as
		\vspace{-2mm}
		\begin{equation}\label{eqn:noitemp}
		\resizebox{0.27\textwidth}{!}{$T_{mol}\left( f,d\right) = T_{o}\left(1-\exp\left(-k\left(f \right)d_{r} \right)  \right)$}  
		\end{equation}
		\begin{equation}
		\resizebox{0.27\textwidth}{!}{$S_{N}\left( f,d_{r}\right)  = S_{N_{B}}\left(f,d_{r} \right)+S_{N_{P}}\left(f,d_{r} \right) $} 
		\end{equation}
		\begin{equation}\label{eqn:mnm1}
		\resizebox{0.43\textwidth}{!}{$S_{N_{B}}(f, d_{r}) = \lim\limits_{d_{r} \rightarrow \infty} k_{B} T_{0}\left( 1-\exp\left( -k\left(f \right)d_{r} \right) \right) \left( \frac{c_{0}}{\sqrt{4\pi}f_{0}}\right)^{2}$} 
		\end{equation}
		\begin{equation}\label{eqn:mnm2}
		\resizebox{0.43\textwidth}{!}{$S_{N_{P}}\left(f,d_{r} \right) = S_{P}\left( f\right)\left( 1-\exp\left( -k\left(f \right)d_{r} \right) \right) \left( \frac{c_{0}}{4\pi d_{r} f_{0}}\right)^{2} $} 
		\end{equation}
		where $k_{B}$ is the Boltzmann constant, $T_{0}$ is the room temperature and $S_{P}\left( f\right) $ represents p.s.d. of transmitted pulse. 
			\begin{figure}
				\vspace{-4mm}
		\centering
		\includegraphics[width= \columnwidth, height = 1.25cm]{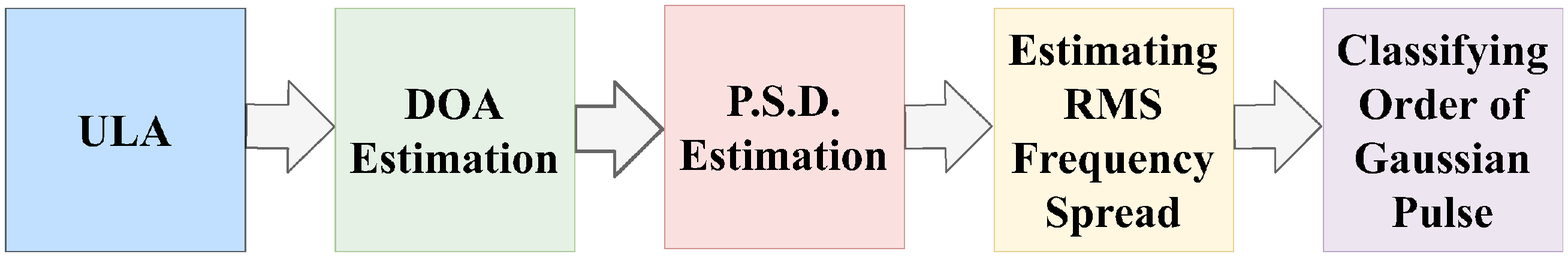}
		\vspace{-5mm}
		\caption{RMS frequency spread based Order classification of Gaussian pulse.}
		\vspace{-5mm}
		\label{fig:BLK_DIAG}
	\end{figure}
\vspace{-3mm}
		\section{System Model}
		\vspace{-0.1cm}
		The proposed system model for estimating the order of the Gaussian pulse is shown in Fig. \ref{fig:BLK_DIAG}. The benefit of using ULA in the system model is, along with the classification of the order of the Gaussian pulse, even the nanosensor node that transmitted the pulse can be localized by estimating its direction of arrival (DOA). The estimation of DOA using wideband multiple signal classification algorithm is studied in \cite{bn}. Further, estimating the center frequency using spectral centroid is attempted but only for a fixed derivative order, i.e., order classification was not investigated \cite{bn1,bn2}. The steps for localizing and estimating the order of Gaussian is summarized as follows\\
	$\bullet$ Localize the nanosensor node that transmitted a higher order Gaussian pulse by estimating its DOA.\\
	$\bullet$ Estimate the p.s.d. of the received Gaussian pulse.\\
	$\bullet$ Estimate the order of the Gaussian pulse by computing RMS frequency spread from estimated p.s.d.
	
\begin{figure}
			\vspace{-2mm}
			\centering
			\subfigure{
				\includegraphics[width=0.48\columnwidth, height = 3cm]{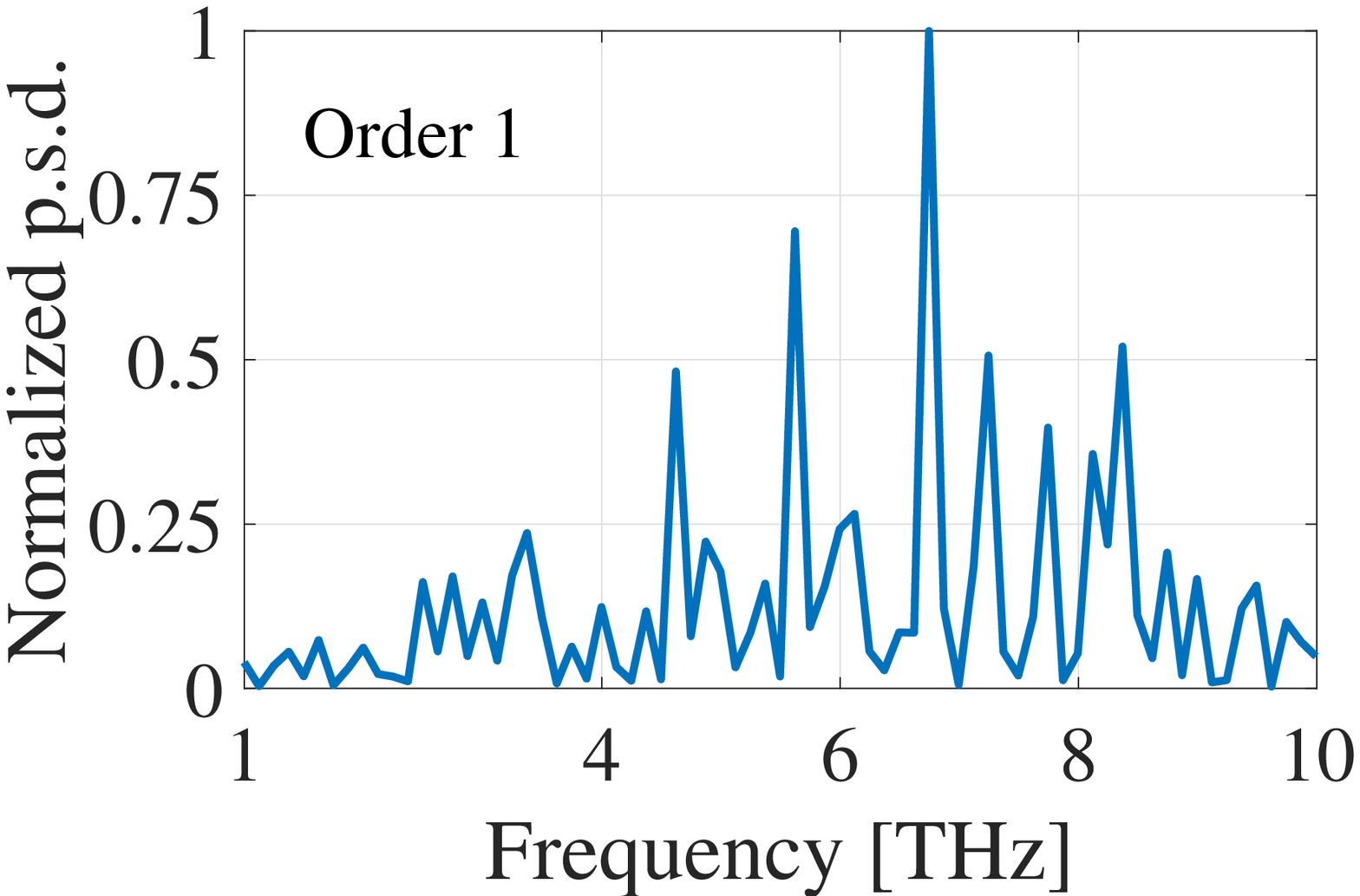}
			}
			\hspace{-6mm}
			\subfigure{
				\includegraphics[width=0.48\columnwidth, height = 3cm]{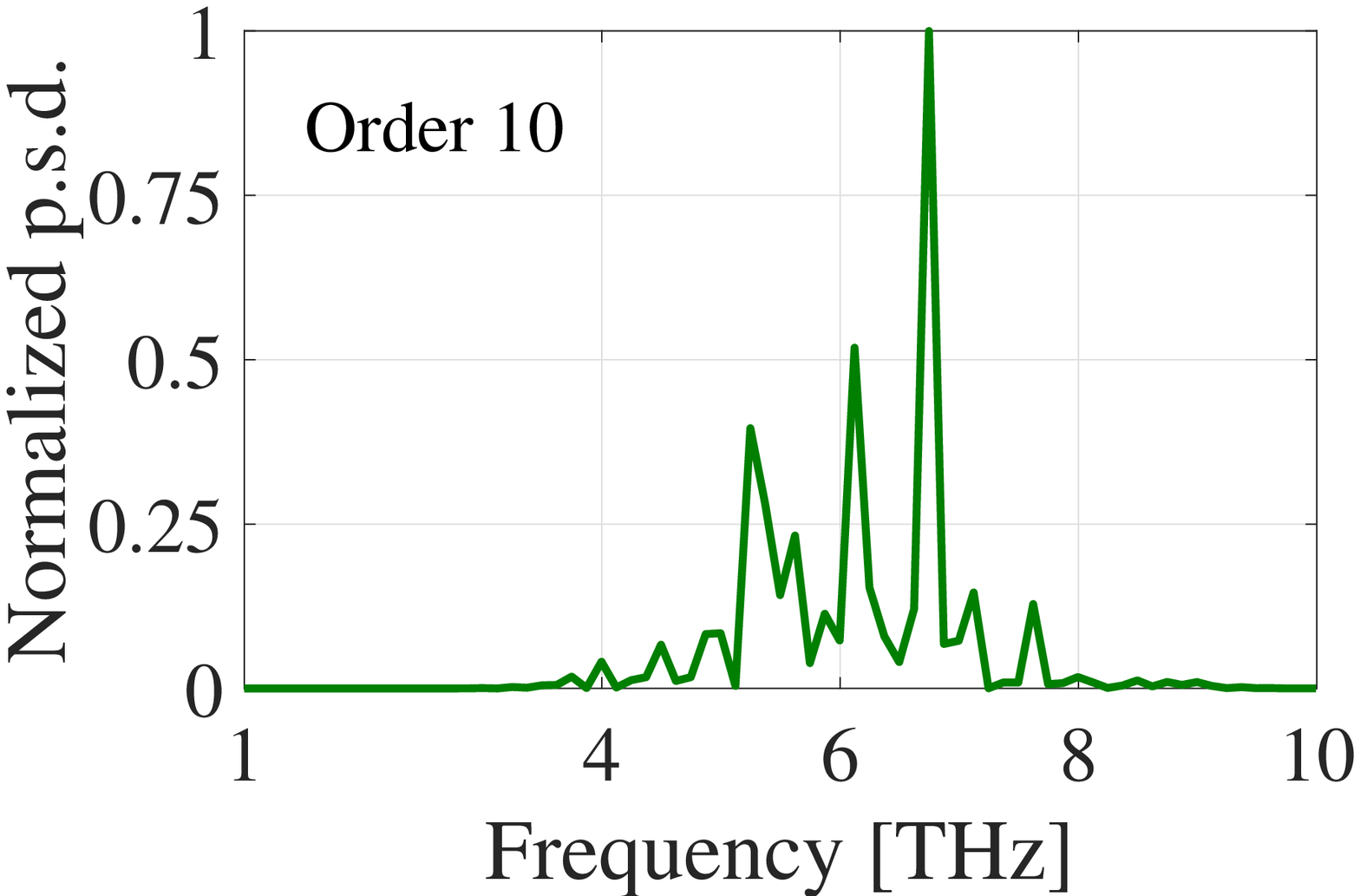}
			}
		\vspace{-2mm}
			\caption{Estimated p.s.d. of order 1 and 10 Gaussian pulse with 6 THz center frequency, 8 ps snapshot observation duration and path length of 1 cm.}
			\label{fig:RMS_SPR_FIG}
			\vspace{-7mm}
		\end{figure}

	We now describe the proposed solution for estimating the order of  Gaussian pulse with a known center frequency.  From \eqref{eq:cent_freq} it is observed that for a fixed center frequency, $\sigma$ value increases with an increase in the order of the Gaussian pulse. Hence at a particular center frequency, lower order Gaussian pulses have wider half power bandwidth compared to higher order Gaussian pulses as observed in Table \ref{tab:TABLE1}. In Table \ref{tab:TABLE1}, $f_{l}$ and $f_{h}$ represents lower and upper half power frequencies and $B_{3\:dB}$ represents half power bandwidth. Hence at a particular known center frequency, using the estimated p.s.d., the order of Gaussian pulse can be identified by locating the half power frequencies (3 dB bandwidth). Fig. \ref{fig:RMS_SPR_FIG} shows the estimated p.s.d. at 6 THz center frequency for $1^{\text{th}}$ and $10^{\text{th}}$ order Gaussian pulse at a path length of 1 cm for single simulation trail. It is observed in Fig. \ref{fig:RMS_SPR_FIG} that, the estimated p.s.d. is no longer Gaussian shaped and it is difficult to locate the half power frequencies. A possible explanation for this outcome is due to molecular absorption loss and molecular absorption noise affecting the propagating pulse in the terahertz channel. This absorption loss and noise depends on the molecular resonance peak of the terahertz channel. But it can be observed in Fig. \ref{fig:RMS_SPR_FIG} that, depending on half power bandwidth the spreading of the p.s.d. around the center frequency will be different. Lower order Gaussian pulses will have larger p.s.d. spread as compared to higher order Gaussian pulses. Hence, we propose the metric RMS frequency spread $\Gamma_{n,f_{c}}$ to identify the order of Gaussian pulse and is defined as     	
	\vspace{-2mm}
		\begin{equation}
		\resizebox{0.5\columnwidth}{!}{$\Gamma_{n,f_{c}} = \sqrt{\frac{\int\limits_{B} \left ( f-f_{c} \right )^{2}S_{n}\left ( f \right ) df}{\int\limits_{B}S_{n}\left ( f \right ) df}}$}
		\vspace{-3mm}
		\end{equation}
		 where $B$ represents the bandwidth of the terahertz channel, $f$ is the frequency and $S_{n}\left ( f \right )$ is the p.s.d. of the Gaussian pulse. The RMS frequency spread in THz for different higher order Gaussian pulses is shown in Table \ref{tab:TABLE1}. From Table \ref{tab:TABLE1} it is observed that for a particular center frequency, RMS frequency spread decreases with an increase in the order of the Gaussian pulse. This decrease in RMS frequency spread is due to decrease in half power bandwidth for higher order Gaussian pulses. Further, the RMS frequency spread decreases rapidly till seventh order Gaussian pulse and thereafter from eight order Gaussian the decrease in RMS frequency spread is very slow. Hence different higher order Gaussian pulses with sufficiently large order difference will have a large difference between their RMS frequency spread values. Therefore in this paper we investigate the order classification performance using $1^{\text{th}}$, $4^{\text{th}}$ and $10^{\text{th}}$ order Gaussian pulses.\\
		Since the p.s.d. of the received Gaussian pulse will be estimated at discrete frequency values, the estimate of RMS frequency spread $\Gamma_{\hat{S}_{n}\left( f\right) }$ is defined as
		\vspace{-1mm}
		\begin{equation}
		\resizebox{0.5\columnwidth}{!}{$\Gamma_{\hat{S}_{n}\left( f\right) } = \sqrt{\frac{\sum\limits_{b = 1}^{L} \left ( f_{b}-f_{c} \right )^{2}\hat{S}_{n}\left ( f_{b} \right )}{\sum\limits_{b=1}^{L}\hat{S}_{n}\left ( f_{b} \right )}}$}
		\end{equation}
		where $\hat{S}_{n}\left ( f \right )$ is the estimated p.s.d.. Using $\Gamma_{\hat{S}_{n}\left( f\right) }$, the order of Gaussian pulse is classified according to the following rule
		\begin{equation}\label{eq:ordest}
		\begin{split}
		\hat{\Gamma}_{\hat{n},\hat{f}_{c}} &= \Gamma_{n_{i},\hat{f}_{c}},\\
		&\text{if}\: \left |\Gamma_{\hat{S}_{n}\left( f\right) }-\Gamma_{n_{i},f_{c}}\right |\leq \left | \Gamma_{\hat{S}_{n}\left( f\right) }-\Gamma_{n_{j},\hat{f}_{c}} \right |\: \forall \: j\neq i
		\end{split}
		\vspace{-1mm}
		\end{equation} 
		The estimation of p.s.d. of the received Gaussian pulse requires the DOA estimate of the nanosensor node. The procedure to estimate the DOA of the nanosensor node is described in the following subsection.
		\vspace{-2mm}
		\subsection{DOA Estimation}
		\vspace{-1mm}
		The DOA estimate of Nanosensor node is obtained using a ULA with $N$ elements. The spacing between antenna elements in ULA is $d_{s}$ m. The path length between ULA and nanosensor device is represented as $d_{r}$. The received wideband higher order Gaussian pulse at the output of the $i$th element in ULA is represented as \cite{b9}.  
		\begin{equation}\label{eq:datam}
		y_{i}\left(t, d_{r}\right) =  p_{n}\left(t - \tau_{i} \right) * h\left(t,d_{r} \right) + v_{i}\left( t,d_{r}\right)   
		\end{equation}
		\begin{equation}\label{eq:delay_eq}
		\tau_{i} = \left(i -1\right)d_{s}\sin\left( \theta\right) /c 
		\end{equation}
		where $h\left(t,d_{r}\right) $ is the terahertz channel impulse response between ULA and nanosensor device in $\theta$ direction. $v_{i}$ represents molecular absorption noise created between element $i$ of ULA and nanosensor device. $\tau_{i}$ represents the time delay for the pulse arriving at $i^{\text{th}}$ element of ULA with respect to a reference element. The time delay $\tau_{i}$ is measured with respect to a reference element of ULA which is located at the origin. When the signal received at the output of the ULA is observed for sufficiently large time interval $\Delta T$ (significantly longer than the propagation time across ULA), the Fourier representation of  \eqref{eq:datam} is given as\cite{b13}
		\vspace{-1mm}
		\begin{IEEEeqnarray}{lll}
			\resizebox{0.41\textwidth}{!}{$Y_{i}\left( f_{b},d_{r}\right) =  e^{-j2\pi f_{b} \tau_{i}} P_{n}\left( f_{b}\right) H\left( f_{b},d_{r}\right) + V_{i,}\left( f_{b},d_{r}\right)$} \\ \nonumber
			\hspace{5cm}\text{for}\; b = 0,\cdots, L
			\vspace{-2mm}
		\end{IEEEeqnarray}
		where $f_{b} $ is the frequency bin, $P_{n}\left( f_{b}\right) $, $H\left( f_{b},d_{r}\right) $, and $V_{i,}\left( f_{b},d_{r}\right)$ are Fourier coefficients of Gaussian pulse, terahertz channel impulse response and molecular absorption noise respectively. Further the output of array is observed for $K$ non-overlapping time interval $\Delta T$ and Fourier coefficients are computed for each time interval. Here, $\Delta T$ is called as snapshot observation duration. For DOA estimation and order classification using \textit{single} pulse, the value of $K$ is set to 1. The number of frequency bins, $L$, in each snapshot is given as
		\begin{equation}\label{eq:timbw}
		L = \lfloor B \cdot \Delta T \rfloor +1
		\end{equation}
		where $\lfloor \cdot  \rfloor$ is the floor operator, $B$ is the terahertz channel bandwidth. From \eqref{eq:timbw} it is observed that, the number of frequency bins $L$ increases with an increase in bandwidth $B$ and snapshot observation duration $\Delta T$. Note that a larger number of frequency bins gives better p.s.d. estimate, which in turn can help increase the order classification accuracy. It will be shown through simulation results (Section V) that, larger snapshot observation duration yields better order classification accuracy. Now, the Fourier coefficients at frequency $f_{b}$ across $N$ sensors for the $K$ number of frequency snapshots is represented in matrix form as 
		\vspace{-1.5mm}
		\begin{equation}\label{eq:datafm_vec}
		\resizebox{0.4\textwidth}{!}{$\boldsymbol{Y}\left(f_{b},d_{r} \right) =  H\left( f_{b},d_{r}\right) \boldsymbol{a}\left(f_{b},\theta \right) \boldsymbol{P}_{n}\left(f_{b} \right) + \boldsymbol{V}\left( f_{b},d_{r}\right)$} 
		\end{equation}
		where $\boldsymbol{Y}\left(f_{b},d_{r} \right) \in \mathbb{C}^{N \times K}$, $\boldsymbol{V}\left( f_{b},d_{r}\right)\in \mathbb{C}^{N \times K}$ and\\ $\boldsymbol{P}_{n}\left(f_{b} \right)\overset{\Delta}{=}\left[ {P}_{n1}\left(f_{b} \right),\cdots,{P}_{nK}\left(f_{b} \right)\right] $.\\$\boldsymbol{a}\left(f_{b},\theta \right) =\left[1, e^{-j2\pi f_{b} \tau_{1}}, \cdots, e^{-j2\pi f_{b} \tau_{N}}\right]^{T} $ is the array manifold vector. The covariance matrix $\boldsymbol{R_{Y}}\left(f_{b},d_{r} \right)$ of $\boldsymbol{Y}\left(f_{b},d_{r} \right)$ is given as 
		\vspace{-1mm}
		\begin{equation}\label{eq:covmat}
		\resizebox{0.6\columnwidth}{!}{$\boldsymbol{R_{Y}}\left(f_{b},d_{r} \right)  = \mathbb{E}\left[ \boldsymbol{Y}\left(f_{b},d_{r} \right)\boldsymbol{Y}\left(f_{b},d_{r} \right)^{H}\right]$} 
		\vspace{-1mm}
		\end{equation}
		where $\left(\cdot \right) ^{H}$ denotes conjugate transpose and $\mathbb{E}\left[ \cdot\right] $ represents expectation. For small path length, where the molecular noise temperature is low, \eqref{eq:covmat} can be simplified as 
		\begin{equation}\label{eq:stdeqn}
		\resizebox{0.875\columnwidth}{0.04\columnwidth}{$\boldsymbol{R_{Y}}\left(f_{b},d_{r} \right)   =	\left|P_{n}\left( f_{b}\right) \right|^{2} \left|H\left( f_{b},d_{r}\right) \right|^{2} \boldsymbol{a}\left( f_{b},\theta\right)\boldsymbol{a}\left( f_{b},\theta\right)^{H} +\sigma^{2}\left( f_{b},d_{r}\right)  \boldsymbol{I}_{N}$}
		\end{equation}
		In \eqref{eq:stdeqn},  $\boldsymbol{I}_{N}$ is the identity matrix of size $N\times N$ and the term \resizebox{0.5\columnwidth}{!}{$\mathbb{E}\left[ \boldsymbol{V}\left( f_{b},d_{r}\right)\boldsymbol{V}^{H}\left( f_{b},d_{r}\right) \right] =\sigma^{2}\left( f_{b},d_{r}\right)$} is the noise variance around narrow frequency sub-band centered at frequency $f_{b}$. Eqn. \eqref{eq:stdeqn} is same as the covariance matrix at the output of ULA assuming noise to be independent of Gaussian pulses emitted by nanosensor devices. $\sigma^{2}\left( f_{b},d_{r}\right)$ is computed as 
		\begin{equation}
		\resizebox{0.45\columnwidth}{!}{$\sigma^{2}\left( f_{b},d_{r}\right) = \int S_{N}(f_{b},d_{r}) df$}
		\vspace{-1mm}
		\end{equation}
		In this paper, incoherent multiple signal classification (IMUSIC) DOA estimation method is used for localizing nanosensor devices which transmits wideband higher order Gaussian pulses. The IMUSIC algorithm can perform DOA estimation even with single pulse or frequency snapshot is because of low molecular absorption noise for path length below 0.5 m \cite{b13_T}.  In IMUSIC DOA estimation technique, narrowband MUSIC DOA estimation technique is independently applied to each $L$ number of frequency bins. The IMUSIC wideband DOA estimation technique is given as \cite{b13_IM}
		\begin{equation}\label{eq:IMUS}
		\resizebox{0.43\textwidth}{!}{$P_{\text{IMUSIC}}( \hat{\theta}, d_{r})  = \sum\limits_{l=1}\limits^{L}\frac{\boldsymbol{a}^{H}\left(f_{b},\theta \right)\boldsymbol{a}\left(f_{b},\theta \right) }{\boldsymbol{a}^{H}\left(f_{b},\theta \right)\boldsymbol{E}_{n}\left(f_{b},d_{r} \right)\boldsymbol{E}_{n}^{H}\left(f_{b},d_{r} \right)\boldsymbol{a}\left(f_{b},\theta \right)} $}
		\end{equation}	
		where $\boldsymbol{E}_{n}\left(f_{b},d_{r} \right)$  is the  noise eigenvector matrix which is obtained from eigen value decomposition of $\boldsymbol{R_{Y}}\left(f_{b},d_{r} \right) $.  Eqn. \eqref{eq:IMUS} is called as IMUSIC spectrum and it is observed that, the quality of DOA estimate depends on communication distance between nanosensor device and ULA. The DOA estimate from IMUSIC spectrum is estimated as 
		\vspace{-2.25mm}
		\begin{equation}\label{eq:tht_est}
		\resizebox{0.55\columnwidth}{!}{$\hat{\theta} \left( d_{r}\right) = \argmax_{\boldsymbol{\theta}}\left[ P_{\text{IMUSIC}}( \hat{\theta}, d_{r})\right] $}
		\end{equation}
		\vspace{-1mm}
		Further, the received covariance matrix at each frequency bin $f_{b}$ is estimated as
		\vspace{-1mm}
		\begin{equation}
		\resizebox{0.6\columnwidth}{!}{$\boldsymbol{\hat{R}_{Y}}\left(f_{b},d_{r} \right) = \frac{1}{K}\boldsymbol{Y}\left(f_{b},d_{r} \right) \boldsymbol{Y}^{H}\left(f_{b},d_{r} \right)$}
		\end{equation}
		\subsection{Estimation of p.s.d.}
		\vspace{-1mm}
		We now describe the procedure for estimating the p.s.d. of the received Gaussian pulse from the estimated DOA. From Fig. \ref{fig:TEMP_GRAP} it is observed that the effect of molecular absorption noise is negligible for path length below 0.5 m. Using this assumption, the noise in \eqref{eq:stdeqn} is neglected and the estimated p.s.d. is given as   
		\vspace{-2mm}
		\begin{equation}\label{eqn:psd_est}
		\resizebox{0.7\columnwidth}{!}{$\hat{S}_{n}\left(f_{b} \right) = \left( \boldsymbol{a} \left(f_{b},\hat{\theta} \right) \right)^{\dagger} \hat{\boldsymbol{R}}_{\boldsymbol{Y}}\left(f_{b}.d_{r} \right)\left( \left( \hat{\boldsymbol{a}} \left(f_{b},\hat{\theta} \right) \right)^{H}\right)^{\dagger}$}  
		\vspace{-2mm}
		\end{equation}
		where $\left( \cdot\right) ^{\dagger}$ represents pseudo inverse operator and $\boldsymbol{a} (f_{b},\hat{\theta})$ is the array steering vector computed using DOA estimate $\hat{\theta}$. 
		\vspace{-2mm}
		\section{Simulation Results}
		\vspace{-1mm}
		In this section, simulation results are presented to analyze the effect of snapshot observation duration and number of antenna elements in ULA on order classification accuracy of a single transmitted Gaussian pulse. Since the half power bandwidth for a given higher order Gaussian pulse is wider at a higher center frequency as compared to lower center frequency, the order classification accuracy is analyzed for $1^{\text{th}}$, $4^{\text{th}}$, and $10^{\text{th}}$ order Gaussian pulses for different center frequencies at 3 THz and 6 THz. The RMS frequency spread estimation algorithm as explained in the previous section is used to classify the order of Gaussian pulse is implemented using MATLAB 2014b. 
	
		In the simulation terahertz channel frequency band is considered from 1 THz to 10 THz. The high-resolution transmission molecular absorption (HITRAN) database \cite{b14} is used to obtain the molecular absorption coefficient $k\left( f\right) $ of the terahertz channel for standard summer air with $1.86 \%$ concentration of water vapor. The distance between consecutive antenna elements is half the wavelength $\lambda_{min}$ of frequency 10 THz, that is $d_{s} = \SI{15}{\micro\meter}$ to avoid spatial aliasing. The transmitting nanosensor device is assumed to be located in the far-field region of ULA with DOA $15.7125^{\circ}$. The power of the higher order Gaussian pulses in the simulation is considered as 1 \si{\micro}W. The minimum snapshot observation time interval is selected as $\Delta T= 2\;\text{ps}$, which is slightly larger than the total pulse duration of highest order Gaussian with lower center frequency ($n\;=\;10\;\text{and}\;f_{c}\;=\;3\;\text{THz}$).
		
		The accuracy of order classification for a particular order is defined as the true positive rate (TPR), which is obtained as the ratio of the number of correct classifications divided by the total number of pulse transmissions simulated for that order. The \textit{average} TPR is computed by taking the average of TPRs corresponding to $1^{\text{th}}$, $4^{\text{th}}$, and $10^{\text{th}}$ order Gaussian pulses. In the simulation, the total number of single pulse transmissions is considered as 200. 
		\vspace{-3mm}
		\subsection{Impact of Center Frequency}
\begin{figure}
			\centering
			\subfigure{
				\includegraphics[width=0.48\columnwidth, height = 3cm]{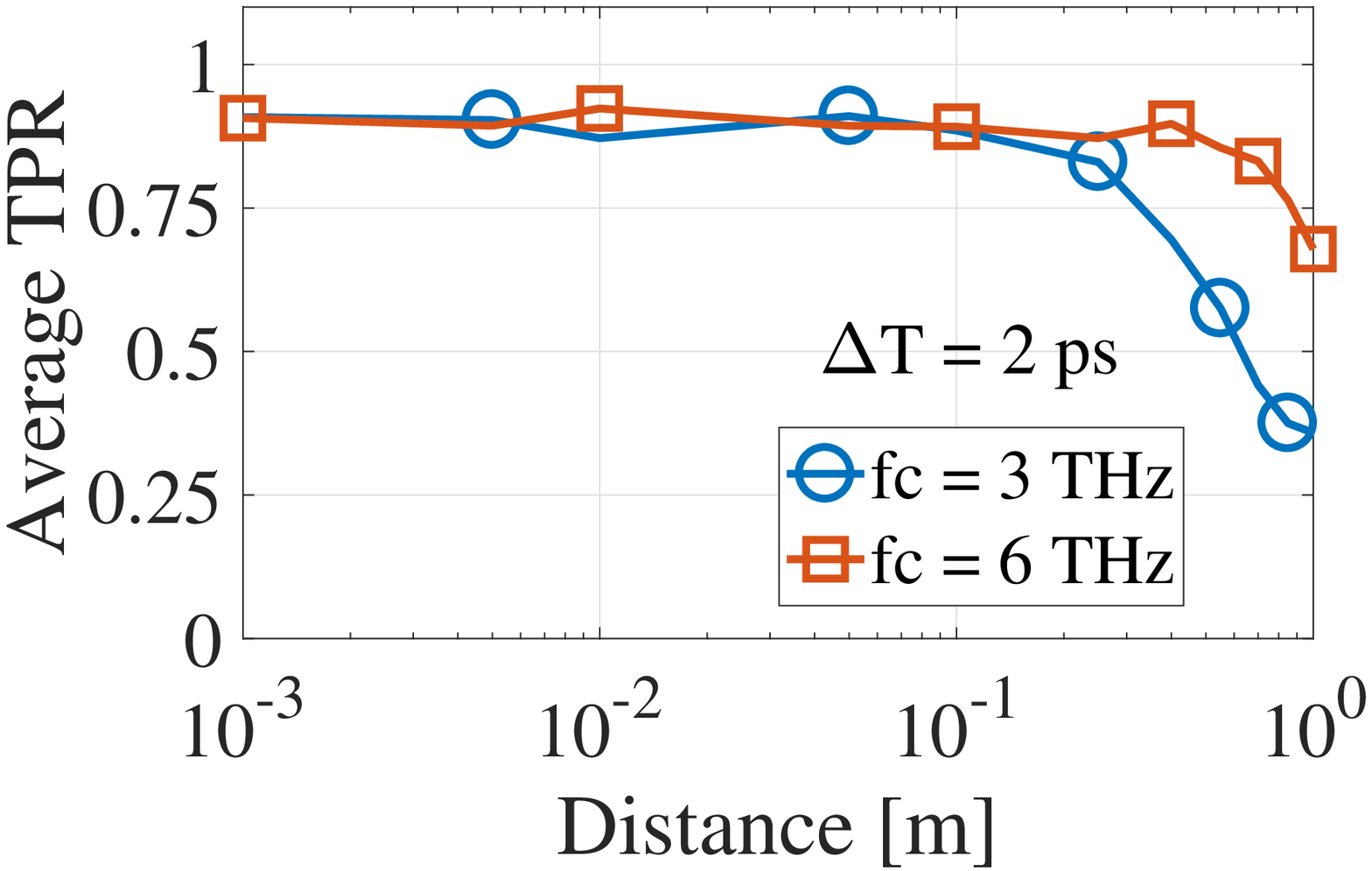}
			}
		\vspace{-2mm}
			\hspace{-6mm}
			\subfigure{
				\includegraphics[width=0.48\columnwidth, height = 3cm]{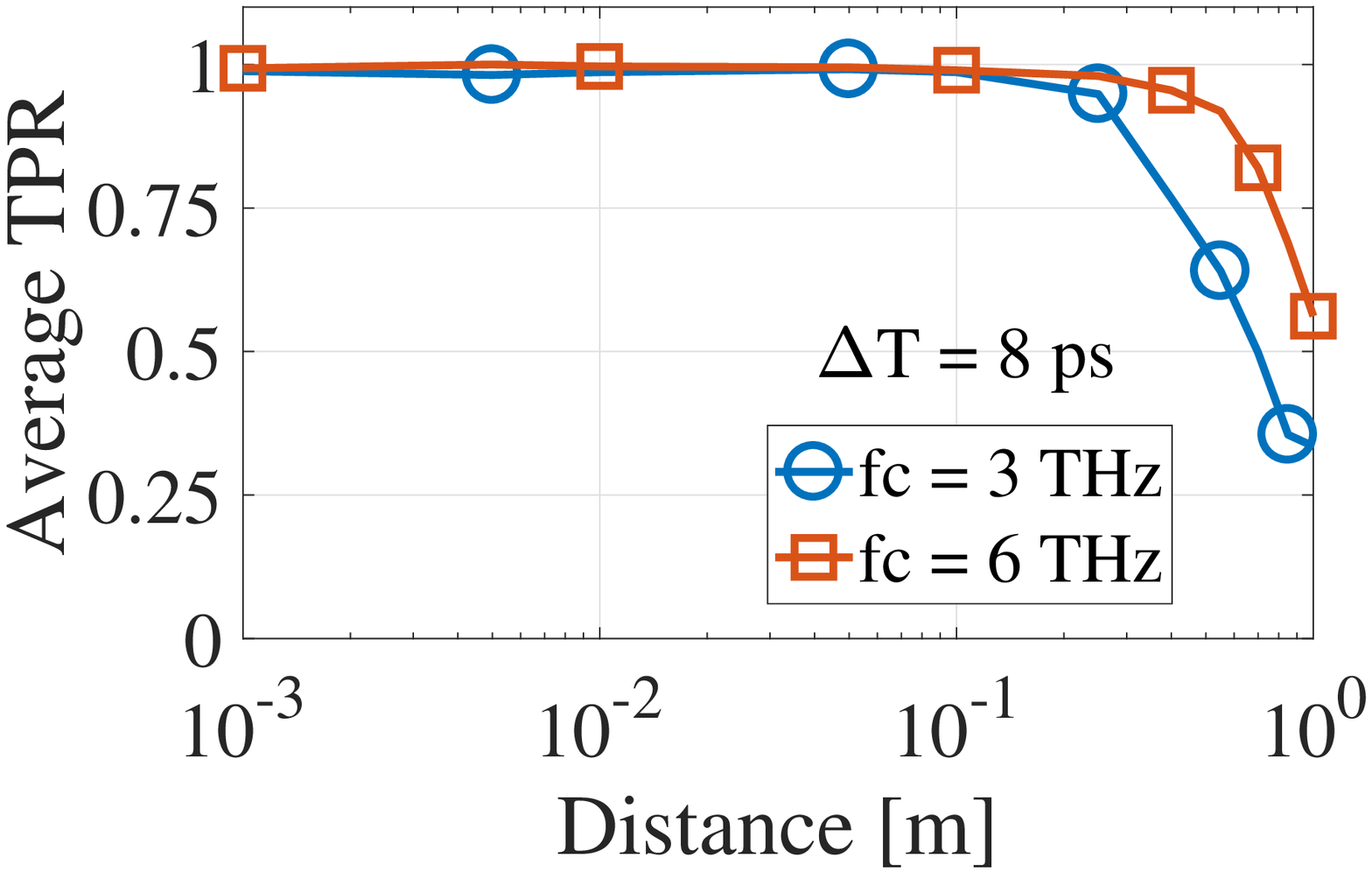}
			}
		    \vspace{-2mm}
			\caption{Average TPR (Normalized) as a function of path length for 3 THz and 6 THz center frequency and 8 antenna elements.}
			\label{fig:AVG_TPR_SS}
			\vspace{-3mm}
		\end{figure}

		\vspace{-1mm}
	Fig. \ref{fig:AVG_TPR_SS} shows the order classification accuracy of the Gaussian pulse as a function of path length for center frequencies 3 THz and 6 THz. We observe that Gaussian pulses focusing their energy at 6THz  outperforms 3 THz pulses when path length is stretched beyond 50 cm. A possible explanation for this outcome is due to wider half power bandwidth for Gaussian pulses at 6 THz center frequency as compared to Gaussian pulses with 3 THz center frequency. With wider half power bandwidth, more number of frequency bins with significant p.s.d. values are included within the half power bandwidth providing an improved estimate of RMS frequency spread, which in turn improves order classification.
		\begin{figure}
		\centering
		\subfigure{\includegraphics[width=0.95\columnwidth, height =3.3cm]{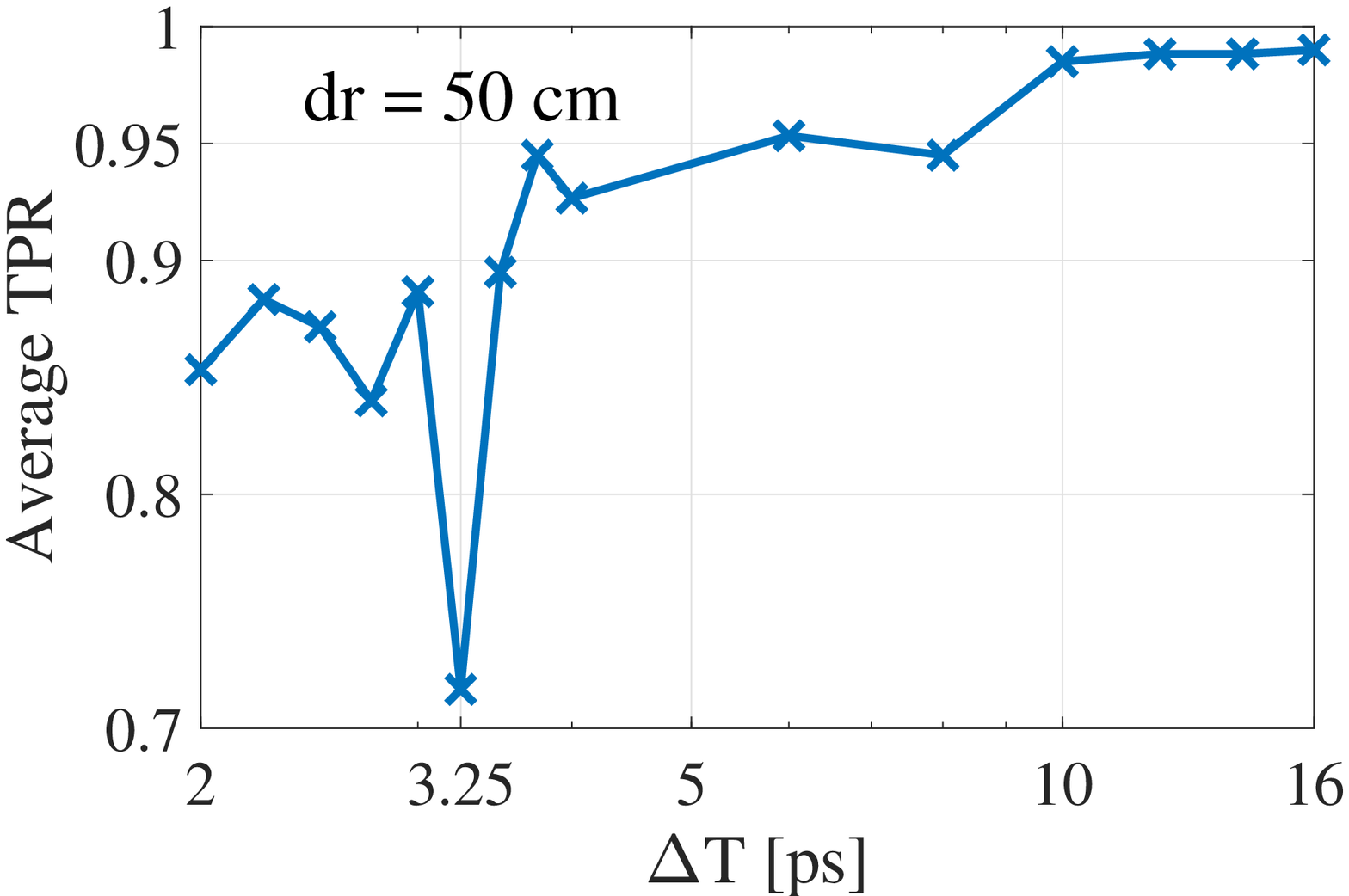}}
		\subfigure{\includegraphics[width=0.95\columnwidth,height=3.3cm]{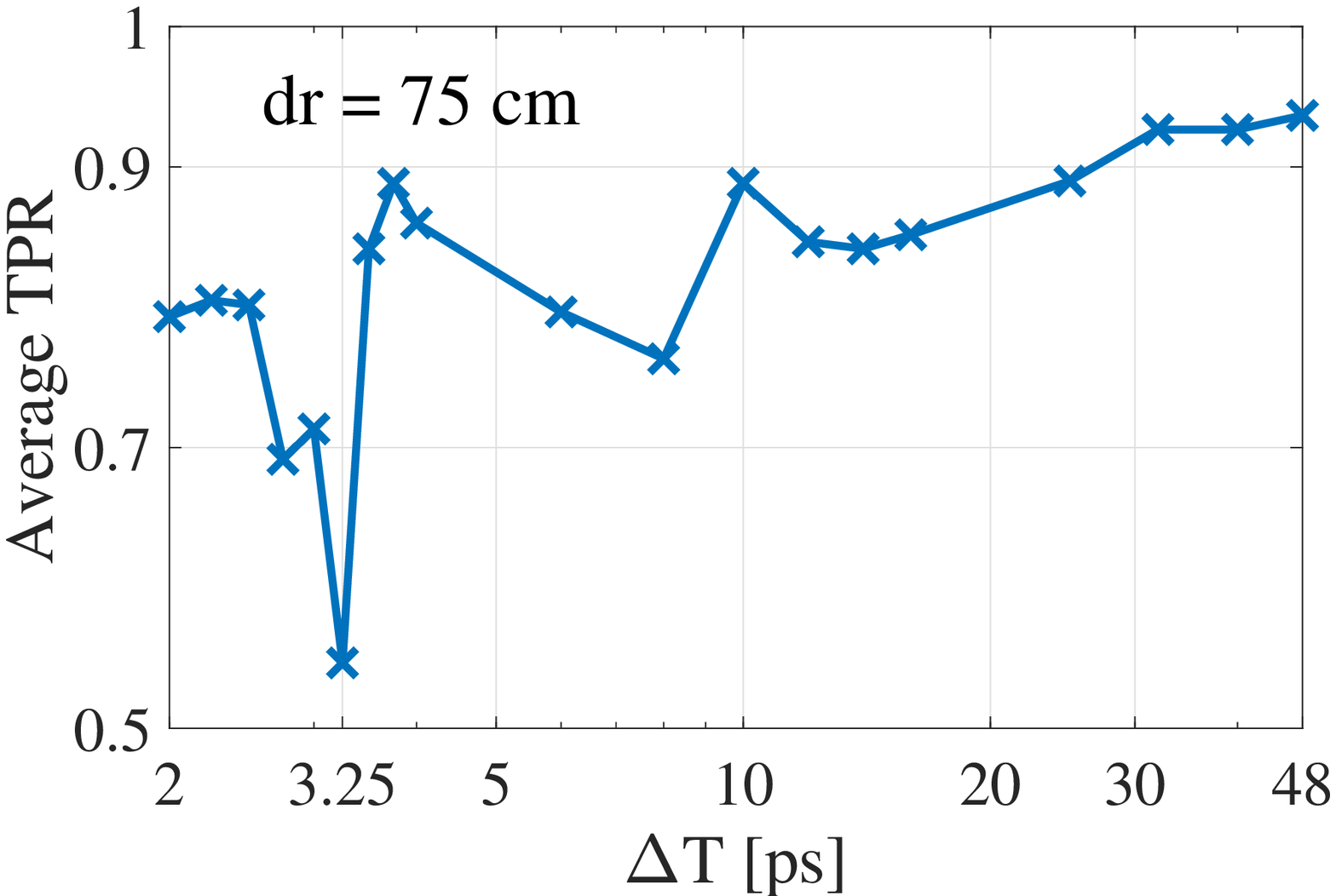}}
		\vspace{-3mm}
		\caption{Average TPR as a function of snapshot observation interval for Gaussian pulses with 6 THz center frequency and at path lengths of 50 cm and 75 cm. The number of antenna elements in ULA is 8.}
		\label{fig:OBS_INT}
		\vspace{-7mm}
	\end{figure}
\vspace{-2mm}
		\subsection{Impact of Snapshot Observation Interval}
\begin{figure}
			\centering
			\subfigure{
				\includegraphics[width=\columnwidth, height = 3.75cm]{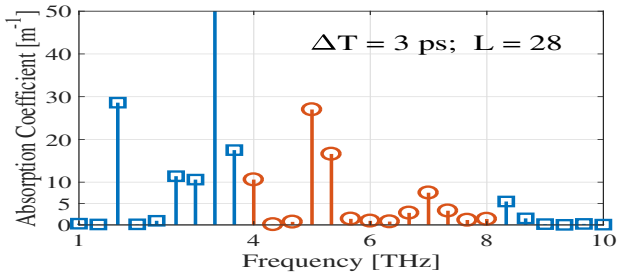}
			\vspace{-7mm}
		}
			\subfigure{
				\includegraphics[width=\columnwidth, height = 3.75cm]{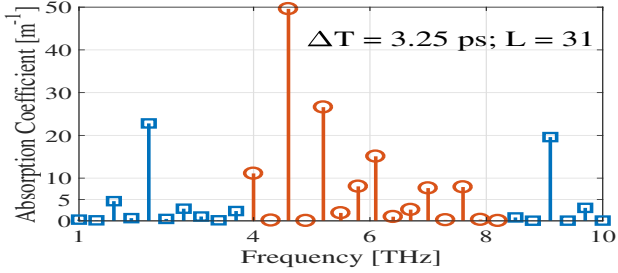}
			}
		    \vspace{-4mm}
			\caption{Molecular absorption coefficient versus frequency bin $f_{b}$ values for snapshot observation time of 3 ps and 3.25 ps. Red and blue colors represent, respectively, the molecular resonance peaks inside and outside of the 4-8 THz band (3 dB bandwidth).}
			\label{fig:freq_bin}
		\end{figure}

		\vspace{-1mm}
		The dependence of order estimation accuracy of the Gaussian pulse on snapshot observation duration is investigated in Fig. \ref{fig:OBS_INT}. We make the following observations. \textbf{1.)} The order classification accuracy increases with increase in duration of snapshot observation time. The reason for the improvement in classification accuracy for higher snapshot observation interval is due to the availability of the higher number of frequency bins, which helps improve p.s.d. estimation. \textbf{2.)} The order classification accuracy at 50 cm is better than that at 75 cm, which is due to the increased molecular absorption noise at longer distances. For path length of 50 cm, the order estimation accuracy is close to 99\% for snapshot observation duration greater than 10 ps, whereas for 75 cm, 48 ps observation is required to achieve only 93\% classification accuracy. \textbf{3.)} It is observed that classification accuracy improvement saturates at 10ps for 50 cm and 25 ps for 75 cm. A possible explanation for this outcome is because the estimated p.s.d. from the \textit{single} snapshot (\textit{single} pulse) is inherently inaccurate, which cannot be overcome without increasing the number of snapshots beyond one \cite{b15}. However, in this paper, we consider detecting the derivative order of a \textit{single} pulse, which may be transmitted once in a while by a sensor to update its status. As such, we only focus on exploring the limit of what can be achieved with a \textit{single} snapshot. \textbf{4.)} Average TPR does not increase \textit{monotonically} with increasing snapshot duration. For example, average TPR is poor for snapshot observation duration of 3.25 ps as compared to 3 ps. A possible explanation for this outcome is explained with the help of Fig. \ref{fig:freq_bin}, which shows the molecular absorption coefficient values at different frequency bin values $f_{b}$. We observe that the snapshot observation duration of 3.25 ps has more molecular resonance peaks near 6 THz frequency as compared to 3 ps. Note that a large number of resonance peaks yields noisy estimate of p.s.d., which makes it harder to classify the order of the pulse.
	\vspace{-1.5mm}
		\subsection{Impact of Number of Antenna elements in ULA}
\begin{figure}
			\vspace{-5mm}
			\centering
			\includegraphics[width=0.95\columnwidth, height =3.5cm]{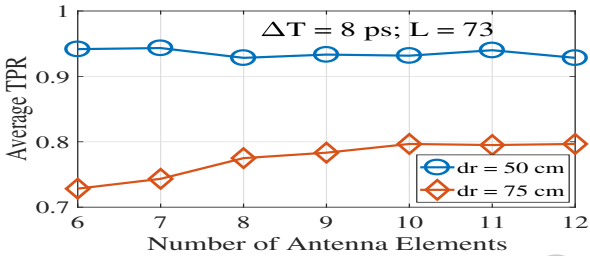}
			\vspace{-3mm}
			\caption{Average TPR as a function of the number of antenna elements in ULA. The center frequency of the Gaussian pulse is 6 THz.}
			\label{fig:NUM_ULA}
			\vspace{-6mm}
		\end{figure}

		\vspace{-0.75mm}
		 Fig. \ref{fig:NUM_ULA} shows order classification accuracy of the Gaussian pulse for varying number of antenna elements in ULA. It is observed that there is no significant improvement in the average TPR with increasing number of antenna elements in ULA. A possible explanation for this outcome is that, the p.s.d. estimate obtained using single pulse is inherently inaccurate and hence the order classification accuracy cannot be improved with an increase in antenna elements in ULA. 
		\vspace{-3mm}
		\section{Conclusion}
		\vspace{-1.5mm}
		We have proposed RMS frequency spread to classify the order of Gaussian pulse using ULA antenna. For a single snapshot, we have analyzed the order classification accuracy for different snapshot observation durations and for varying number of antenna elements in ULA. Our investigation shows that, for a single snapshot, order classification accuracy for Gaussian pulses depends on the duration of snapshot observation rather than on the number of antenna elements in ULA. Large path length requires large snapshot observation duration to achieve order classification accuracy greater than 90\%. Snapshot duration of 16 ps can achieve 99\% classification accuracy for 50 cm distance, while 48 ps is required to achieve 93\% accuracy at 75 cm. To achieve higher classification accuracies at longer distances, in our future work, we will explore more advanced methods, such as compressive sampling for p.s.d. estimation, and machine learning for order classifications.
		\vspace{-3mm}
		\section*{Acknowledgment}
		\vspace{-1mm}
		The proposed work is supported by SERB, GOI under order no. SB/S3/EECE/210/2016
		
	\end{document}